\documentstyle[12pt]{article} 
\def \be{\begin{equation}} 
\def \ee{\end{equation}} 
\begin{document} \begin{flushright} TIFR/TH/97-37 \\ 
10 July 1997
\end{flushright} 
\bibliographystyle{unsrt} 
\vskip0.5cm
\baselineskip=.8cm 
\begin{center} 
{ \LARGE 
\bf Two-particle correlations and $B^0\bar B^0$ mixing }  \\ 
[7mm] { 
\bf G.V. Dass$^a$ and K.V.L. Sarma}$^{b,*}$\\ 
$^a${ \it Department of Physics, Indian
Institute of Technology, Powai, Mumbai, 400 076, India }\\ 
\vskip0.3cm
$^b${ \it Tata Institute of Fundamental Research, Homi Bhabha Road,
Mumbai, 400 005, India } \\ 
[10mm] 
\end{center}

\bigskip 
------------------------------------------------------------------------\\
{\bf Abstract}

\bigskip

\baselineskip=.8cm 

We study the EPR correlation implied by the entangled wavefunction of
the $B^0\overline{B}^0$ pair created by the $\Upsilon $(4S) resonance.
The analysis uses the basis provided by the mass eigenstates $B_1,B_2$
rather than the flavour states $B^0, \overline{B}^0$. Data on the
inclusive dilepton charge ratio are close to the expectation of
quantum mechanics, but nearly 8 standard deviations away from that of
complete decoherence.

\bigskip

\bigskip

---------------------------------------------------------------------------\\
\vfill

$^*$~E-mail: kvls@theory.tifr.res.in;~~fax: 091 22 215 2110 

\newpage 
\baselineskip=.8cm 

\newpage

{\bf 1. Introduction}

Bertlmann and Grimus \cite{BG} have made an interesting use of the
dilepton-decay data of the neutral bottom meson pair emitted by the
$\Upsilon$(4S). They focussed on a test of the two-particle
correlation characteristic of quantum mechanics, as manifested over
macroscopic distances ($\sim 10^{-3}$ cm). The tested feature of
quantum mechanics is the consequence of interference between the two
parts of the $ C$-odd wavefunction
\be |\psi_{0}\rangle =
\frac{1}{\sqrt {2}} | B^0 \overline{B}^0 - \overline{B}^0 B^0
\rangle~, 
\label{wf} 
\ee 
where $B^0$ is the pseudoscalar bottom meson with $\overline{b}d$ as
its valence quark constituents. The usual framework of the
Weisskopf-Wigner approximation is assumed so that well-defined
mixtures of the states $B^0$ and $\overline{B}^0$ 
propagate {\it independently}
as $B_1$ and $B_2$ with masses ($m_j$) and inverse-lifetimes ($\Gamma
_j$), ($j=1,2$):
\begin{eqnarray} | B_1
\rangle &=& p |B^0 \rangle + q | \overline{B}^0 \rangle \nonumber \\
[2mm] | B_2\rangle &=& p |B^0 \rangle - q | \overline{B}^0\rangle ~.
\label{b12} 
\end{eqnarray} 
Here $p$ and $q$ are complex constants obeying the normalization
condition $|p|^2+|q|^2=1$, and invariance under $CPT$ is assumed.

The analysis hinges mainly on two experimental ingredients: (i) the
measured ratio of like-sign and unlike-sign dileptons in the chain
decay
\[ \Upsilon(4S) \rightarrow B^0\overline{B}^0 \rightarrow (\ell
^{+,-} +~\cdots ~) + (\ell ^{+,- } +~\cdots ~) \] 
where $\ell $ stands for $e$ or $\mu $, and (ii) the mass difference
$\Delta m= m_2-m_1$ which is extracted from data on $B^0
\overline{B}^0$ oscillations. The strategy adopted by Bertlmann and
Grimus is to confront the data with the standard formula which is
modified by introducing a factor $(1-\zeta )$ in the interference
term. The results, as expected on the basis of other tests of quantum
mechanics to date, are quite consistent with the prediction of quantum
mechanics, i.e., with $\zeta =0 $. The value $\zeta =1$, corresponding
to the extreme case of adding probabilities for the time-dependent
decays of the two parts $B^0 \overline{B}^0$ and $\overline{B}^0 B^0$
of the wavefunction of Eq. (\ref{wf}), is excluded. The $\Delta
B=\Delta Q$ rule which provides a unique relationship between the
lepton charge and the parent bottom flavour (as in the Standard Model)
enters the analysis.

Correlation in the ($B^0,\overline{B}^0$) basis is, however, different
from that in the ($B_1,B_2$) basis. This is readily seen by expressing
the initial state of Eq. (\ref{wf}) as 
\be |\psi _0\rangle =
\frac{1}{2\sqrt {2}pq} |B_2 B_1 -B_1 B_2 \rangle , \label{wfp} 
\ee 
and allowing the states (of definite momenta) to evolve to the 
respective decay instants.\footnote{ 
The assumption of $CPT$-invariance, whereby
the constants $p$ and $q$ used in defining $|B_2\rangle $ become the
same as those used in $|B_1\rangle $, is not necessary for the
equivalence of the states given in Eqs. (\ref{wf}) 
and (\ref{wfp}).}
The interference arising from decays of the parts $| B^0
\overline{B}^0 \rangle $ and $|\overline{B}^0B^0 \rangle $ of
Eq. (\ref{wf}) will not be the same as that from decays of the parts
$|B_2B_1 \rangle $ and $|B_1B_2\rangle $ of Eq. (\ref{wfp}), in spite
of the similar looking forms of Eqs. (\ref{wf}) and (\ref{wfp}). The
reason is that a $|B_{1,2}\rangle $ remains a $|B_{1,2}\rangle $ on
time-development, but a $|B^0\rangle $ picks up a
$|\overline{B}^0\rangle $ component (similarly,
$|\overline{B}^0\rangle $ picks up a $|{B}^0\rangle $ component) on
time-development; the interference term in the ($B_1,B_2$) basis turns
out to be comparatively simple. Independently of this simplicity, it
is necessary to subject every correlation implied by quantum-mechanics
to experimental checks; the analysis in one basis is as important as
that in any other. Analysis in the ($B_1, B_2 $) basis is the
motivation behind the present note. In the following we multiply the
interference between the decay amplitudes of the two parts of the
state in Eq. (\ref{wfp}) by a parameter $E$, so that $E=1$ corresponds
to the validity of quantum mechanics and $E=0$ corresponds to the
incoherent addition of decay probabilities. Obviously, $E$ and
$(1-\zeta )$ are not identical because the interference terms they
modify are very different.

{\bf 2. Correlation in the two bases}

We begin with the time-evolved state starting from Eq. (\ref{wfp}) 
\be
|\psi (t',t)\rangle = \frac{1}{2\sqrt {2}pq}~ [~\theta_2 (t') 
\theta_1(t) |
B_2 B_1 \rangle - \theta_1(t') \theta_2(t) | B_1 B_2\rangle ~].  
\ee
Here $t'$ is the proper time of the first beon (say, the one moving 
into
the left hemisphere in the $\Upsilon $ frame), and $t$ of the second
beon; $\theta $'s are the evolution amplitudes \[ \theta_j(t) =\exp
\left[ (-im_j -
              {\Gamma_j\over 2})t \right]~;~~(j=1,2).
\]
The double-decay distribution for the first beon to decay into a
channel denoted by $\beta$ and the second to decay into another
channel $\alpha$ is given (apart from an irrelevant overall constant)
by
\be 
D \left[ \beta (t'), \alpha (t)\right] = 
\frac{1}{8|pq|^2}
\left| \theta_2(t') \theta_1(t) A_{2\beta }A_{1\alpha }
- \theta_1(t') \theta_2(t) A_{1\beta }A_{2\alpha } \right|^2 
                                                      \label{DD} 
\ee
where 
\be 
A_{1\alpha }=p A_\alpha + q \overline{A}_{ \alpha }~;~
A_{2\alpha }= pA_\alpha -q\overline{A}_{\alpha } ~;
\ee
\be
A_\alpha = \langle \alpha |T | B^0 \rangle ~,~ \overline{A}_{
\alpha }= \langle {\alpha } |T| \overline{B}^0 \rangle,
                                                \label{notation} 
\ee 
with similar definitions for the amplitudes corresponding to channel
$\beta$.  

We insert a time-independent, real parameter $E$ (the `EPR factor') in
the interference between the two parts of the state in Eq.
(\ref{wfp}), so that $E=1$ leads to the result of quantum
mechanics. The case of $E=0$ refers to complete decoherence, and is
called the `Furry hypothesis'. With the parameter $E$ inserted, the
time dependence reads as
\begin{eqnarray}
D_E\left[\beta (t'),\alpha (t)\right]& =& \frac{1}{8|pq|^2} ~
\big\{~ 
\left| \theta_2(t') \theta_1(t) A_{2\beta }A_{1\alpha }
\right|^2 + 
\left| \theta_1(t') \theta_2(t) A_{1\beta }A_{2\alpha } 
\right|^2  \nonumber \\
&~&~~~~~~
-2E~{\rm Re}~[ \theta_2^*(t')\theta_1^*(t) 
A_{2\beta }^*A_{1\alpha }^*~ \theta_1(t') \theta_2(t)
A_{1\beta }A_{2\alpha } ]   ~
\big\}.                                   \label{ee}      
\end{eqnarray}

In order to contrast this formula with the one obtained in the ($B^0,
\overline{B}^0$) basis \cite{BG}, we list the ``physical''
states $|B^0(t)\rangle $ and $|\overline{B}^0(t)\rangle $ which evolve
from the initial $|B^0\rangle $ and $|\overline{B}^0\rangle $ states:
\begin{eqnarray}
|B^0(t)\rangle  &=& \theta_+(t)|B^0\rangle + \frac{q}{p}~
\theta_-(t) |\overline{B}^0\rangle ~,\\
|\overline{B}^0(t)\rangle &=&\frac{p}{q}~\theta_-(t) |B^0\rangle +
\theta_+(t) |\overline{B}^0\rangle ~,
\end{eqnarray}
where  
\[ \theta_{\pm}(t) \equiv \frac{1}{2}[\theta _1(t) \pm 
\theta _2(t)]. 
\] 
The corresponding decay amplitudes which are time-dependent will
be denoted by
\begin{eqnarray} 
{\cal A}_{\alpha }(t) &\equiv & \langle \alpha |T| B^0(t)
\rangle ~=~\theta_+(t) A_{\alpha } + \frac{q}{p}~
\theta_-(t) \overline{A}_{\alpha }~,\\
\overline{{\cal A}}_{\alpha }(t)&\equiv & \langle \alpha |T|
\overline{B}^0(t)\rangle ~=~ \frac{p}{q}~
\theta_-(t) A_{\alpha }+\theta_+(t) \overline{A}_{\alpha }~,
\end{eqnarray}
and similar quantities for the channel $\beta $.

In the basis provided by $B^0$ and $\overline{B}^0$ therefore, the 
double-decay probability becomes 
\be 
D \left[ \beta (t'), \alpha (t)\right] = \frac{1}{2} 
\left| {\cal A}_{\beta }(t') \overline{{\cal A}}_{\alpha }(t) - 
\overline{{\cal A}}_{\beta }(t'){\cal A}_{\alpha }(t) \right|^2 ~.
\label{DD1} 
\ee
This expression is modified \cite{BG} by introducing the
decoherence parameter $\zeta $ (following Ref. \cite{EBE}) as
\begin{eqnarray}
D_{\zeta }\left[ \beta (t'), \alpha (t) \right] 
&=& \frac{1}{2} \big\{~ 
| {\cal A}_{\beta }(t') \overline{{\cal A}}_{
\alpha }(t) |^2 + |\overline{{\cal A}}_{
\beta }(t'){\cal A}_{\alpha }(t) |^2 
\nonumber \\
&~&~~~~~-2~(1 - \zeta )~{\rm Re} [  
{\cal A}_{\beta }^*(t')\overline{{\cal A}}_{\alpha }^*(t)~ 
\overline{{\cal A}}_{\beta }(t'){\cal A}_{\alpha }(t) ]~   
\big\}~.                                            \label{ze}  
\end{eqnarray}

We emphasize that while Eqs. (\ref{DD}) and (\ref{DD1}) are the same,
their modified versions Eqs. (\ref{ee}) and (\ref{ze}) are not. Hence
there is no relation between the parameters $E$ and $\zeta $, except
when quantum mechanics is valid which corresponds to $E=1 $ and $\zeta
=0$. It is also to be noted that in the extreme limit of complete
decoherence corresponding to $E=0 $ and $\zeta =1$, the two bases give
different results.

{\it Dileptonic charge ratio}: To proceed further, we specialize 
the decay
amplitudes of neutral beons to semileptonic channels which are
labelled by $m$ and $\overline{m}$, as follows: 
\begin{eqnarray} A_m
&=& \langle m_h~ \ell ^+ ~\nu _{\ell }~|T | B^0 \rangle ~,\\
\overline{A}_{ \bar m }
&=& \langle \overline{m}_h~ \ell ^- ~
\bar {\nu _{\ell } }~ |T| \overline{B}^0 \rangle ~; 
\end{eqnarray} 
where $m_h $ denotes a specified hadronic system having a net negative
charge, and $\overline{m}_h $ denotes its $CPT$ conjugate. In order to
construct the fraction $\chi_d$ of inclusive like-sign dileptons, we
first consider the decay rates into exclusive semileptonic channels,
integrate the rates over $t'$ and $t$, and sum over channel indices.
In this way we obtain the leptonic combinations $\ell ^+\ell ^+$,~
$\ell ^-\ell ^-$,~ $(\ell ^+\ell ^-~+~\ell ^-\ell ^+)$. It is
remarkable that even the inclusive time-integrated correlation among
the lepton charges becomes useful \cite{DH} for testing the subtle
effects of interference of a two-particle entangled wavefunction.

Using the formula (\ref{ee}), we express the dilepton ratio $\chi_d$
as
\begin{eqnarray}
\chi_d &\equiv & \frac{R}{1+R }\\
R&=&\frac{ N(\ell ^+\ell ^+) + N(\ell ^-\ell ^- )}
{N(\ell ^+\ell ^-) + N(\ell ^-\ell ^+)   }\\
&=& \frac{1-aE}{1+aE}\cdot 
\frac{ |p/q|^2\sum_{m,n} |A_mA_n|^2
+|q/p|^2 \sum_{\bar m, \bar n }
|\overline{A}_{\bar m}\overline{A}_{\bar n}|^2  }
{\sum_{m,\bar n}|A_m\overline{A}_{\bar n}|^2~ 
+\sum_{\bar m,n} |\overline{A}_{\bar m} A_n |^2~ }~,
\end{eqnarray}
where for the sake of brevity we used
\be  a \equiv \frac{1-y^2}{1+x^2} , \ee
and the mixing parameters $x$ and $y$ are defined as usual:
\be 
x = \frac{m_2-m_1}{\Gamma},~ \ y =
\frac{\Gamma_2-\Gamma_1}{2\Gamma},~ \ \Gamma = \frac{\Gamma_1
+ \Gamma_2}{2}~.  
\ee

We now use a consequence of $CPT$ invariance which states that the
total semileptonic decay widths of the particle and its antiparticle
are equal,
\[ \sum_{m}|A_m|^2 =
\sum_{\bar n}|\overline{A}_{\bar n} |^2 ~,\]
provided we retain the weak Hamiltonian to first order and neglect the
final state interactions due to electroweak forces.
Using this result we obtain the formula for the likesign dilepton
ratio, 
\begin{eqnarray} 
\chi_d & = & { u~(1-aE) \over (1+u) + 
(1-u)~aE} ~,                                      \\ 
u &\equiv & {1\over 2}[~\left| p/q \right|^2 +\left| q/p \right|^2~]~.
\end{eqnarray}
The parameter $E$ can be determined
\be E= \frac{1+x^2}{1-y^2}\cdot \left[1 - \frac{2\chi_d
}{u+(1-u)\chi_d }\right], \label{master}   
\ee
by using the current experimental information on $u$, $\chi_d $, $x $
and $y$.

{\bf 3. Experimental information}

We note that it is adequate to replace the parameter $u$ by unity in
Eq. (\ref{master}): The dilepton charge asymmetry $a_{\ell \ell}$ at
the $\Upsilon $(4S) measured by the CLEO group \cite{CLE} is
\be 
a_{\ell \ell}\equiv \frac{N (\ell ^+ \ell ^+) - N(\ell
^-\ell ^-)}{ N(\ell ^+\ell ^+) +N(\ell ^-\ell ^-)}= (3.1\pm 9.6\pm
3.2) \times 10^{-2} ~. 
\ee 
As this asymmetry is expressible \cite{OKUN} in terms of the mixing
parameters that define the mass eigenstates,
\[ a_{\ell \ell}=\frac{|p|^4-|q|^4}{|p|^4+|q|^4}~, \] 
the experimental number implies that 
\begin{eqnarray} u &=
&\frac{1}{\sqrt{1- a_{\ell \ell}^2} } \simeq 1+ \frac{1}{2} a_{\ell
\ell}^2 \\ &\simeq & 1.0005 \pm 0.0031~.  
\end{eqnarray} 
Consequently it is safe to use
\be  u \simeq 1~,                        \label{uu} \ee 
since the remaining parameters in Eq. (\ref{master}) are known to much
lower (at least by an order of magnitude) precision.

For the fraction $\chi_d$ , we take the average value given by the
ARGUS \cite{ARG} and CLEO \cite{CLE} collaborations (see, the number
called `our average' on p. 506 of Ref. \cite{PDG}),
\be \chi_d =0.156 \pm 0.024 ~. \label{chie} \ee 
This result is based on $\Upsilon $(4S) data; it does not use the
$\Delta m$ from experiments on oscillations following $Z$ decays.

For the value of $x$, we use the mass-difference $\Delta m$ extracted
from the time-dependence of $B^0_d \overline{B }^0_d$ oscillations in
$Z$ decays. It should however be emphasized that every oscillation
experiment has backgrounds peculiar to it and their subtraction
depends on the use of different algorithms. Background estimates and
simulations do use the Standard Model, not to mention the framework
provided by quantum mechanics; e.g., the jet-charge technique is based
on modelling the $b$-quark fragmentation into a jet. Does it mean that
in obtaining the experimental $\Delta m$, the theory we want to test
is already used? The answer is ``most probably not'': The oscillating
$B$'s are created incoherently through the inclusive decays
$Z\rightarrow B_d~+~\cdots ~ $, while our focus is on a correlation
relating to the coherence of a two-particle $C$-odd wavefunction.

Notwithstanding the above remark it is necessary to follow a
conservative approach so that the results do not depend much on
background estimates and related issues. For the present therefore,
one may like to deal with data having a clean sample of identified
$B_d$ decays, as for instance in the data in which the semileptonic
decays of $B_d$ are reconstructed as completely as possible, although
this would entail a considerable loss in statistics.  The OPAL group
\cite{OPAL} reported a sample of 1200 $D^{*\pm }\ell ^{\mp } $
candidate events, of which 778 $\pm $84 were supposed to be arising
from the decays $B_d \rightarrow D^\ast(2010) \ell \nu X $. The
production flavour was determined by the jet charge method.
Multiplying the $\Delta m$ obtained in this experiment by the average
lifetime of the $B_d$ meson \cite{PDG}, we obtain 
\begin{eqnarray}
x_{{\rm OPAL}} &=&(0.539 \pm 0.060 \pm 0.024)~{\rm ps}^{-1}
\cdot (1.56 \pm 0.06)~{\rm ps}~        \\
&=& 0.84\pm 0.11,                             \label{xopal}
\end{eqnarray}
where we have added the statistical and systematic errors in
quadrature.

On substituting the Eqs. (\ref{uu}), (\ref{chie}), and (\ref{xopal})
in Eq. (\ref{master}), we obtain 
\be E = \frac{1.17\pm 0.15}{ 1-y^2}. \label{yeq} \ee 
At present there is no direct experimental information on the
parameter $y$. It is generally surmised that its magnitude cannot
exceed a few per cent mainly because there exists no flavourless
channel into which both $B_d$ and $\overline{B}_d $ decay
dominantly. But fortunately, even values like $|y|\simeq 0.1$
hardly matter, as the correction to $E$ is only quadratic in $y$. We
may therefore set $y=0$ to obtain
\be E = 1.17\pm 0.15.                    \label{result} \ee
This value of $E$ is consistent with $E=1$ dictated by quantum
mechanics, and is about eight standard deviations away from the Furry
hypothesis ($E=0$). 

On the other hand, following Ref. \cite{BG}, we might assume the
validity of quantum mechanics ($E=1$) and determine $y$ from
Eq. (\ref{yeq}); the resulting bound is not restrictive: $ |y|\leq
0.28 $ at 90\% confidence level.

To what value of $\zeta $ does our result in Eq. (\ref{result})
correspond ? For $y=0$ and $u=1$, we express the likesign dilepton
fraction $\chi _d$ in terms of $\zeta $ to get \be \zeta = \left[
2\chi_d - {x^2\over 1+x^2} \right] {(1+x^2)^2 \over x^2} = - 0.42\pm
0.31.  
\ee 
The numerical value, which results from the substitution of Eqs.
(\ref{chie}) and (\ref{xopal}), is comparable to the one
standard-deviation limit $\zeta \leq 0.26 $ of Ref. \cite{BG}. If we
evaluate $E$ from Eq. (\ref{master}) using the data quoted in
Ref. \cite{BG} ($ {\bar x}=0.74\pm 0.05 $ and $\bar \chi_d= 0.159\pm
0.031$ along with $y=0$ and $u=1$), we would obtain $E=1.06\pm 0.11$.

\newpage

{\bf 4. Summary}

In summary, data on the dileptonic decays of the $B^0_d
\overline{B}^0_d$ pair belonging to $\Upsilon $(4S) are analysed for
possible violations of the two-particle correlation which occurs in
quantum-mechanics. This is done in the basis provided by the mass
eigenstates $|B_1\rangle $ and $|B_2\rangle $. The result,
Eq. (\ref{result}), is completely consistent with quantum-mechanics,
but disfavours the hypothesis of complete decoherence. Although our
use of certain consequences of $CPT$ invariance in $B$ decays may be
innocuous, the generality of our result is reduced by (i) the use of
Weisskopf-Wigner approximation, (ii) the use of the $\Delta B=\Delta Q
$ rule, and (iii) the neglect of the $y^2$ dependence.

\end{document}